\documentclass[twocolumn,showpacs,preprintnumbers,amsmath,amssymb,prb,aps]{revtex4-1}
\usepackage{graphicx}
\usepackage{bm}
\usepackage{amsmath}
\usepackage{multirow}
\begin{document}
\title{Mixed ground state in Fe-Ni Invar alloys}
\author{S. S. Acharya$^1$}
\author{V. R. R. Medicherla$^{1}$}
\email[Corresponding author:]{mvramarao1@gmail.com}
\author{Komal Bapna$^2$}
\author{Khadiza Ali$^2$}
\author{Deepnarayan Biswas$^2$}
\author{Rajeev Rawat$^3$}
\author{Kalobaran Maiti$^2$}
\affiliation{$^1$Department of Physics, ITER, Siksha 'O' Anusandhan Deemed to be University, Bhubaneswar 751030, India}
\affiliation{$^2$Department of Condensed Matter Physics and Materials Science, Tata Institute of Fundamental Research, Homi Bhabha Road,Colaba, Mumbai 400005, India}
\affiliation{$^3$UGC-DAE Consortium for Scientific Research, University Campus, Khandwa Road, Indore 452017, India}
\date{\today}

\begin{abstract}
We investigate the ground state properties of Invar alloys via detailed study of the electronic structure of Fe$_{1-x}$Ni$_x$ alloys ($x$ = 0.2, 0.3, 0.4, 0.5, 0.6, 0.7, 0.9) employing $x$-ray photoelectron spectroscopy (XPS). While all the alloys exhibit soft ferromagnetic behavior with Curie temperature much higher than the room temperature, the results for invar alloy, Fe$_{0.6}$Ni$_{0.4}$ exhibit anomalous behavior. Moreover, the magneto-resistance of the Invar alloy becomes highly negative while the end members possess positive magneto-resistance. The core level spectra of the Invar alloy exhibit emergence of a distinct new feature below 20~K while all other Fe-Ni alloys exhibit no temperature dependence down to 10~K. Interestingly, the shallow core level spectra (3$s$, 3$p$) of Fe and Ni of the Invar alloy reveal stronger deviation at low temperatures compared to the deep core levels (2$s$, 2$p$) indicating crystal field effect. It appears that there is a large precipitation of antiferromagnetic $\gamma^\prime$ phase below 20 K possessing low magnetic moment (0.5$\mu_B$) on Fe within the $\alpha$ phase. The discovery of negative magneto-resistance, anomalous magnetization at low temperature and the emergence of unusual new features in the core levels at low temperature provide an evidence of mixed phase in the ground state of Invar alloys.
\end{abstract}

\maketitle
\section{Introduction}

Magnetic and elastic properties of Fe$_{1-x}$Ni$_x$ alloys are strongly influenced by composition, temperature and pressure. The magnetic phase diagram of the alloy system is enormously rich. The alloys exhibit ferromagnetic behaviour at all compositions. The most striking anomaly observed in low Ni concentration fcc alloys (30-45\% Ni) is, no thermal expansion over a wide range of temperatures \cite{was1991,inv1997} around the room temperature and is called Invar anomaly. First-principle calculations of ferromagnetic $\gamma$-Fe, ordered Fe$_3$Ni and fcc Fe$_{1-x}$Ni$_x$ alloys suggest the existence of two magnetic states \cite{wassermann1990,schroeter1995,and1977,roy1977,morn1990,morz1990,mohn1991,ent1993,john1990,aka1993,
john1997,abr1995}. Recent calculations demonstrated that the Invar anomaly can be related to a continuous magnetic phase transition in Fe-Ni alloys involving a family of complex magnetic states \cite{abr2007}. While disorder and vacancies in a system can give rise to interesting magnetism \cite{CaB6}, the magnetism in Invar alloys is believed to be influenced by thermodynamic and elastic properties of the alloys.

The ground state of Fe-Ni Invar alloys was reported to be inhomogeneous using several experimental techniques \cite{groundstate}. For example, M\"{o}ssbauer studies on the Invar alloy, Fe$_{65}$Ni$_{35}$ suggested that about 90\% of Fe atoms are in ferromagnetic state and the remaining 10\% are either in antiferromagnetic state or in frustrated antiferromagnetic state \cite{mossabauer1}. On the other hand, M\"{o}ssabauer studies by Abd-Elmeguid \textit{et al.} suggested that 30\% of Fe atoms in Fe$_{65}$Ni$_{35}$ are in antiferromagnetic state below 30 K and flip their spins favouring ferromagnetic order above 30 K \cite{abd}. Based on AC susceptibility measurements, Miyazaki \textit{et al.} proposed a reentrant spin glass behaviour for Fe$_{1-x}$Ni$_x$ ($0.3<x<0.45$) below 30 K \cite{miyazaki}. The physical properties of Fe-Ni Invar alloys were found to be anomalous at low temperatures and may be associated to the occurrence of mixed ground state \cite{shakti,rancourt1990}. Evidently, the ground state of Fe-Ni Invars is complex and the investigation of low temperature electronic structure of Fe-Ni alloys may shed light in understanding the ground state properties of Fe-Ni Invars.

In this study, we have prepared the whole composition range of Fe$_{1-x}$Ni$_{x}$ alloys from $x$ = 0.0 to 1.0 and investigated their electronic structure using photoemission spectroscopy. While the core level spectra of non-Invar Fe-Ni alloys exhibit expected temperature dependance, the results for the Invar alloy, (Fe$_{0.6}$Ni$_{0.4}$) are anomalous below 20 K suggesting the existence of a mixed phase in the ground state.

\section{Experiment}

Fe$_{1-x}$Ni$_x$ ($x$ = 0.1, 0.2, 0.3, 0.4, 0.5, 0.6, 0.7, 0.8, 0.9) have been prepared by arc melting method under highly pure argon atmosphere. The ingots were melted several times to ensure homogeneity of the sample and were characterized using $x$-ray diffraction (XRD), resistivity and magnetization measurements. The XRD studies indicate a structural transition from bcc to fcc structure as a function of composition \cite{shakti}. Fe$_{0.7}$Ni$_{0.3}$ alloy forms in mixed phase containing both bcc and fcc phases while all other compositions are in single phase. The $x$ = 0.4 alloy falls in the Invar regime. X-ray photoelectron spectroscopy(XPS) measurements have been carried out using PHOIBOS 150 electron Analyzer from Specs GmbH and monochromatic Al $K_\alpha$ radiation (1486.6 eV) as an excitation source at a base pressure of about 10$^{-10}$ Torr. Considering the large intrinsic linewidth of the core level spectra under study, the total energy resolution of the spectrometer was fixed at 0.5 eV in order to achieve good signal to noise ratio without compromising intensities significantly. The Fermi level was derived using a highly pure Ag sample mounted on the same sample holder so that it is in electrical contact with the sample. We tried to cleave the samples using a top-post. However, the effort was not successful as the arc melted samples were very hard. Therefore, the alloy surfaces were cleaned by polishing in ultra high vacuum condition using a fine grained diamond file. Reproducibility of the spectra was ensured after each cycle of surface preparation. The surface cleanliness was ascertained by the absence of O 1$s$ and C 1$s$ core level signals. Temperature variation between 10 K and 300 K was done using an open-cycle helium cryostat, LT-3M from Advanced Research Systems.

\section{Results}

\begin{figure}
  \centering
   \includegraphics[width = 0.9\linewidth]{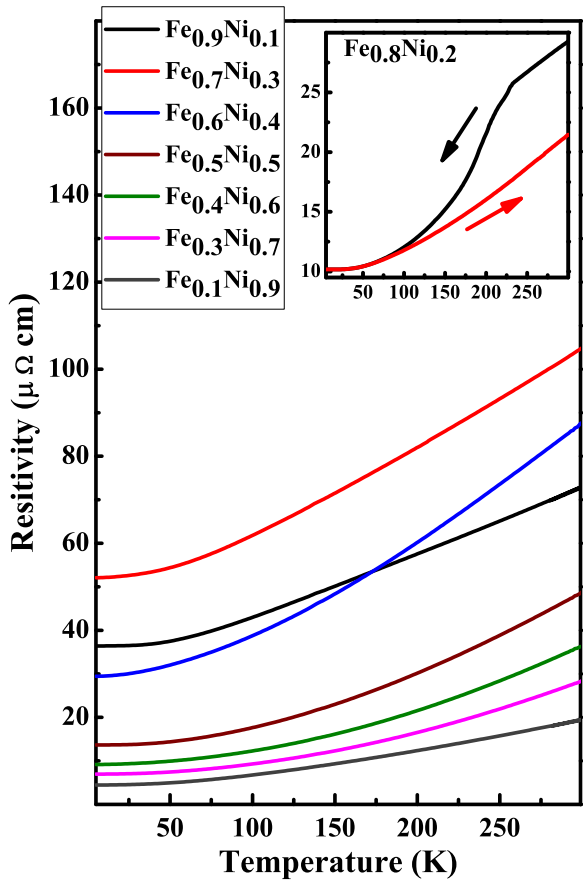}
 \vspace{-2ex}
\caption{The temperature dependance of the resistivity of Fe$_{1-x}$Ni$_x$ alloys for various values of $x$; the values of $x$ is given in the figure legend.}
\label{Fig1-res}
 \vspace{0ex}
\end{figure}

The temperature dependance of resistivity of Fe$_{1-x}$Ni$_x$ alloys is shown in Fig. \ref{Fig1-res}. All the compositions exhibit a typical metallic behavior. The resistivity of the Fe$_{0.8}$Ni$_{0.2}$ alloy, shown in the inset of Fig. \ref{Fig1-res} exhibit hysteresis for heating and cooling cycles due to martensite transition \cite{shakti}; the resistivity curves start to deviate at a temperature of about 55 K, the difference grows with increase in temperature and goes beyond the room temperature. The resistivity of Fe$_{0.7}$Ni$_{0.3}$ alloy containing mixed structural phases is higher than the resistivity of all other compositions although the behavior continues to be metallic in the whole temperature range studied. Such higher resistivity may be attributed to disorder arising from the presence of mixed bcc and fcc phases. The resistivity is found to decrease with the enhancement of Ni concentration, which may be associated to the enhancement of 3$d$ occupation. Moreover, fcc structure is a close packed structure with packing fraction significantly larger than bcc structure along with a larger number of nearest neighbors. Therefore, the bandwidth is expected to be larger in the fcc structure and hence higher degree of itineracy, which might be a reason for decrease in resistivity with increase in Ni concentration.

\begin{figure}
 \vspace{0ex}
  \centering
\includegraphics [width = 0.9\linewidth]{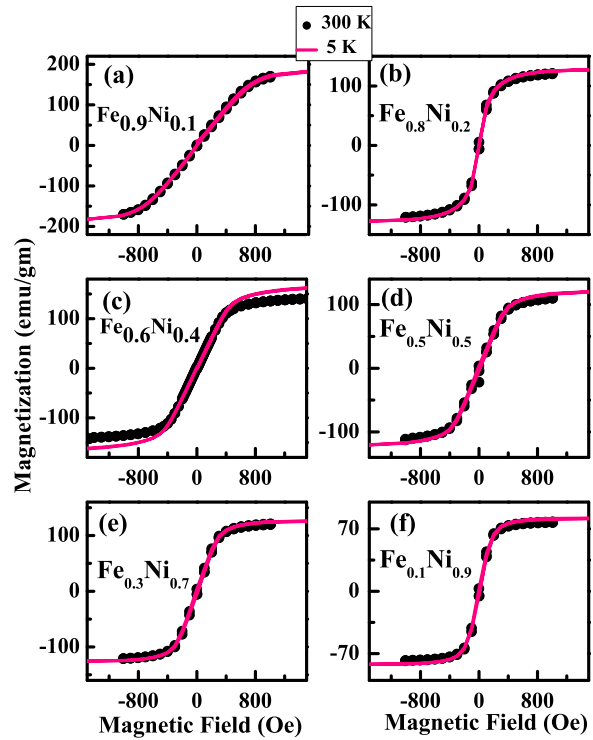}
 \vspace{-2ex}
\caption{The magnetic hysteresis loops of Fe$_{1-x}$Ni$_x$ alloys for various values of $x$. The symbols represent the data at 300 K and the lines are the ones at 5 K.}
\label{Fig2-BHloop}
 \vspace{0ex}
\end{figure}

In Fig. \ref{Fig2-BHloop}, we show the magnetic hysteresis loops of Fe$_{1-x}$Ni$_x$ alloys recorded at 300 K (symbols) and 5 K (lines); the data were superimposed over each other for comparison. All the compositions exhibit hysteresis typical of a soft ferromagnetic behavior. The Fe$_{0.9}$Ni$_{0.1}$ alloy exhibits highest saturation magnetization of about 170 emu/gm even at 300 K. The saturation moment gradually decreases with the increase in Ni concentration, which can be attributed to the smaller Ni moment compared to Fe moment. However, the magnetic interaction appears to increase with the higher Ni content leading to a saturation at lower magnetic field. Gradual change with composition indicates absence of anomalies due to disorder and/or vacancies expected in intermediate compositions \cite{CaB6}. The hysteresis loops recorded at 300 K and 5 K temperatures overlap on each other for all the compositions except for the Fe$_{0.6}$Ni$_{0.4}$ Invar alloy; the saturation moment at 5 K is significantly higher than the value at 300 K. Moreover, hardness of magnetization appears to be larger in this case compared to the compositions on both sides. These observations suggest the presence of different magnetic ground state for Invar alloy compared to other compositions.

\begin{figure}
 \vspace{0ex}
  \centering
\includegraphics [width = 0.9\linewidth]{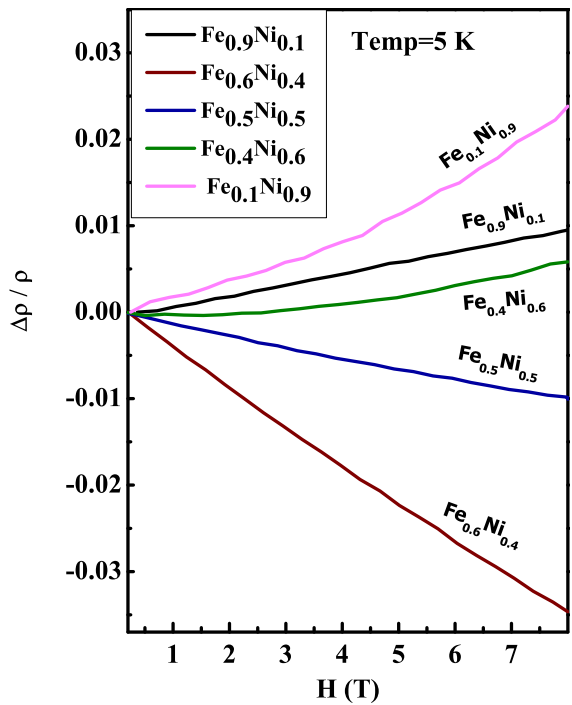}
 \vspace{-2ex}
\caption{The field dependance of the magneto-resistance of Fe$_{1-x}$Ni$_x$ alloys measured at 5 K. The magneto-resistance changes sign with the change in composition. While the end members exhibit positive magneto-resistance, the intermediate compositions show negative value. The magnitude of magneto-resistance is highest for the Invar alloy.}

\label{Fig3-MR}
 \vspace{0ex}
\end{figure}

The external magnetic field dependance of the magneto-resistance (MR) of Fe$_{1-x}$Ni$_x$ alloys measured at 5 K is shown in Fig. \ref{Fig3-MR}. High Fe and high Ni concentration alloys exhibit positive magneto-resistance - a typical behavior expected in metals due to enhancement of resistivity with the application of external magnetic field. The behavior becomes significantly different in the intermediate compositions Fe$_{0.6}$Ni$_{0.4}$ and Fe$_{0.5}$Ni$_{0.5}$ alloys exhibiting negative magneto-resistance, which corresponds to a decrease of resistivity with the application of magnetic field. The results for the composition, Fe$_{0.4}$Ni$_{0.6}$ is close to zero. The magnitude of the magneto-resistance for Fe$_{0.5}$Ni$_{0.5}$ alloy is small while Fe$_{0.6}$Ni$_{0.4}$ Invar alloy exhibits relatively larger value. The negative magneto-resistance is an indicative of the presence of antiferromagnetic interactions within the ferromagnetic matrix. While this is in line with enhancement of hardness in magnetism observed in the hysteresis curves of the system, enhancement of saturation magnetization at low temperature observed in Fig. \ref{Fig2-BHloop} cannot be explained by this behavior. One reason for such behavior can be the reduction of Curie temperature leading to thermal effects more prominent near the room temperature, which is absent in other compositions. Clearly, the ground state of this material is complex.

\begin{figure}
 \vspace{0ex}
  \centering
\includegraphics [width = 0.7\linewidth]{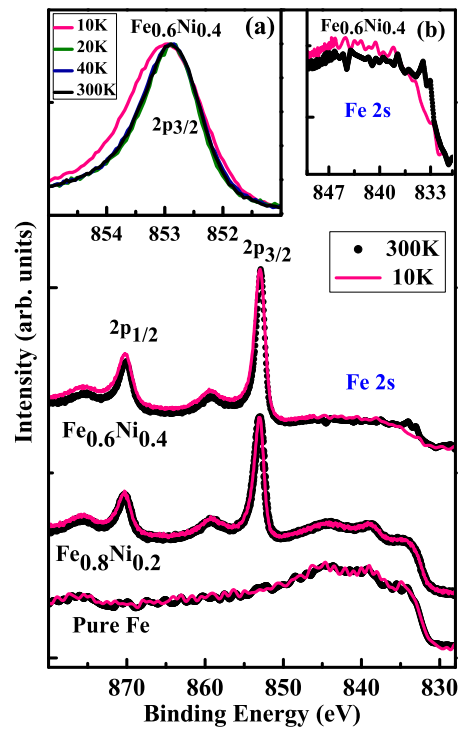}
 \vspace{-2ex}
\caption{Ni 2$p$ and Fe 2$s$ spectra of Fe, Fe$_{0.8}$Ni$_{0.2}$, Fe$_{0.6}$Ni$_{0.4}$ recorded at 300 K (symbols) and 10 K (lines). The spectra at 10 K are overlayed on the spectra at 300 K. Inset (a) shows the Ni 2p$_{3/2}$ spectra of  Fe$_{0.6}$Ni$_{0.4}$ recorded at the temperatures, 10 K, 20 K, 40 K and 300 K. Inset (b) shows the Fe 2$s$ spectra of Fe$_{0.6}$Ni$_{0.4}$ recorded at 300 K and 10 K.}
\label{Fig4-Ni2p}
 \vspace{0ex}
\end{figure}

In Fig. \ref{Fig4-Ni2p}, we show the Ni 2$p$ spectra of Fe$_{0.8}$Ni$_{0.2}$ and Fe$_{0.6}$Ni$_{0.4}$ (Invar) alloys recorded at 300 K and 10 K. On the low binding energy side of Ni~$2p_{3/2}$ peak, a broad Fe 2$s$ feature is observed. The Fe 2$s$ peak of pure Fe is also shown for reference. The main peak in the 2$p$ spectra corresponds to the well screened final state, where the core hole is screened by a conduction electron and the satellite at 6 eV higher binding  energy is due to the poorly screened final state \cite{core-Maiti,raether}. A similar set of main peak and satellite structure is found for the spin-orbit split 2$p_{1/2}$ core excitation as shown in the figure. The spectra recorded at 300 K and 10 K overlap exactly for Fe$_{0.8}$Ni$_{0.2}$ alloy, but the 10 K spectrum of Fe$_{0.6}$Ni$_{0.4}$ Invar alloy looks slightly broader in the 2$p$ region and seems to be narrower in Fe 2$s$ region.

The temperature induced difference in the 2$p_{3/2}$ core level spectra of Fe$_{0.6}$Ni$_{0.4}$ Invar alloy is shown in the inset (a) in Fig. \ref{Fig4-Ni2p}. The 2$p_{3/2}$ peak at 300 K exhibits asymmetry towards higher binding energy side; a typical behavior of the core level spectra in metallic systems due to low energy excitations across the Fermi level along with the core level excitations. The lineshape remains unchanged down to 20 K. However, at 10 K, the spectrum became significantly broad, which is unusual as the phonon contributions becomes less significant at low temperatures and one expects a narrowing instead.

The spectral functions for the Fe 2$s$ excitations in pure Fe and Fe$_{0.8}$Ni$_{0.2}$ samples exhibit three features at 844.9 eV, 838.6 eV and 833.9 eV. The relative intensity and the peak positions of the Fe 2$s$ features in pure Fe and Fe$_{0.8}$Ni$_{0.2}$ alloy collected at 300~K and 10~K are almost identical as evident in the figure. The spectral features in Fe$_{0.6}$Ni$_{0.4}$ Invar alloy are significantly different; distinct structures in the Fe 2$s$ spectra are not visible. Interestingly, the temperature seems to have a profound effect on the low binding energy side of the 2$s$ spectra as shown in the inset (b) of the figure. There is a sharp decrease in intensity at around 833.9 eV at 10 K relative to the intensity in the 300~K spectra. In addition, the intensity around 844 eV has got enhanced.

\begin{figure}
 \vspace{0ex}
   \centering
\includegraphics [scale=0.4]{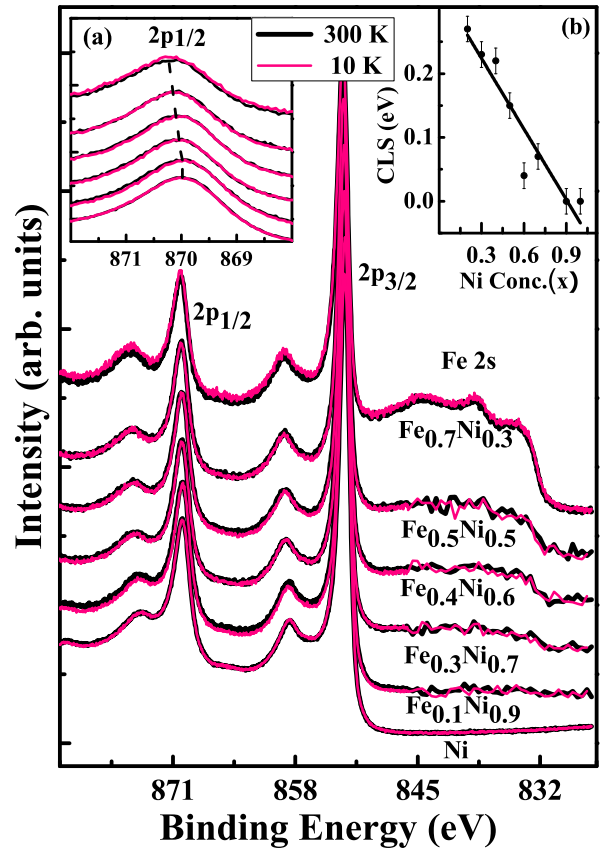}
 \vspace{-2ex}
\caption{Ni 2$p$ and Fe 2$s$ spectra of Fe$_{1-x}$Ni$_x$ alloys for various values of $x$ recorded at 300 K (symbols) and 10 K (lines). The 10 K spectra are superimposed over the spectra taken at 300 K for comparison. Inset (a) shows the binding energy shift in 2p$_{1/2}$ feature with alloy composition; while Inset (b) shows the variation of CLS's with Ni concentration.CLS's follow a linear behaviour with Ni concentration}
\label{Fig5-Ni2pAll}
 \vspace{0ex}
\end{figure}

Ni 2$p$ spectra along with the Fe 2$s$ photoemission signals for various compositions, $x$ = 0.3, 0.5, 0.6, 0.7, 0.9 and 1.0 are shown in Fig. \ref{Fig5-Ni2pAll}. The spectra recorded at 300~K and 10~K temperatures are superimposed over each other and exhibit almost identical spectral lineshape at both the temperatures indicating absence of temperature induced effect in these cases. Fe 2$s$ intensities gradually reduce with the increase in Ni concentration as expected. The Ni 2$p$ feature exhibits a change with the change in Ni concentration; the main peak shifts towards lower binding energies and the satellite appear to shift towards the main peak with the increase in Ni concentration.

The core level shift (CLS) relative to the feature in pure Ni is shown in the inset of the figure. The derivation of the CLS from 2$p_{3/2}$ feature is ambiguous as Fe 2$s$ features are very close in energy to 2p$_{3/2}$ features. Therefore, we have used 2p$_{1/2}$ peak to derive the CLS's. The CLS's are positive for all alloys and exhibit a linear dependence with negative slope on increasing Ni concentration. Among the cases studied, the maximum shift of 0.27~eV was observed for $x$ = 0.2 alloy (lowest Ni concentration). Observed CLS's for each alloy were found to be the same at 300 K and 10 K temperatures.

The core level shift towards higher binding energies with the increase in relative Fe concentration indicates an effective increase in Madelung potential at the Ni site with the enhancement of Fe neighbours around Ni \cite{YBCO-Hike}; presumably linked to effective higher positive charge at the Fe sites. Interestingly, along with the core level shift, the relative separation of the satellite peak and the main peak also enhances with the enhancement of Fe content suggesting significant modification in electron interaction parameters in the final state of photoemission \cite{core-Maiti}.

\begin{figure}
 \vspace{-2ex}
   \centering
\includegraphics [scale=0.25]{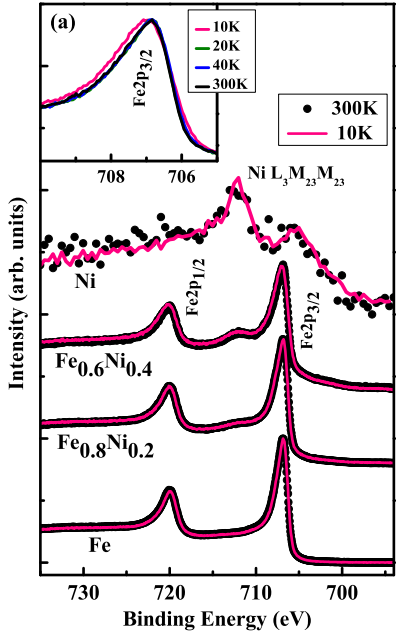}
 \vspace{-2ex}
\caption{Fe 2$p$ spectra of Fe$_{1-x}$Ni$_x$ for $x$ = 0.0, 0.2, 0.4, 1.0 recorded at 10 K (lines) and 300 K (symbols). The spectra at 10 K are overlayed on the spectra at 300 K. Ni $L_3M_{23}M_{23}$ Auger feature of pure Ni is shown for reference. Inset shows Fe 2$p_{3/2}$ signal of Fe$_{0.6}$Ni$_{0.4}$ recorded at temperatures, 10 K, 20 K, 40 K and 300 K.}
\label{Fig6-Fe2p}
 \vspace{-2ex}
\end{figure}

Fe 2$p$ spectra of pure Fe, Fe$_{0.8}$Ni$_{0.2}$ and Fe$_{0.6}$Ni$_{0.4}$ are shown in Fig. \ref{Fig6-Fe2p}. We also show the $L_3M_{23}M_{23}$ Auger feature of pure Ni metal in the figure to identify the features in the Fe 2$p$ spectra as the Fe 2$p$ features in the alloys appear in similar energy regime. The spectra recorded at 10~K are overlayed on the spectra at 300~K for comparison. The spectra of pure Fe and Fe$_{0.8}$Ni$_{0.2}$ samples match well for both the temperatures, whereas the spectra of Fe$_{0.6}$Ni$_{0.4}$ Invar alloy recorded at 10~K is broader compared to that recorded at 300~K - a similar behavior observed in the Ni 2$p$ core level spectra. The anomalous cooling induced broadening of the 2$p_{3/2}$ peak is shown in the inset of the figure. It is evident that the spectral lineshape at and above 20 K is identical to that at room temperature exhibiting an asymmetry towards higher binding energies typical of a metal. The anomaly is observed in the 10 K spectrum, which is significantly broader.

\begin{figure}
 \vspace{-2ex}
   \centering
\includegraphics [scale=0.4]{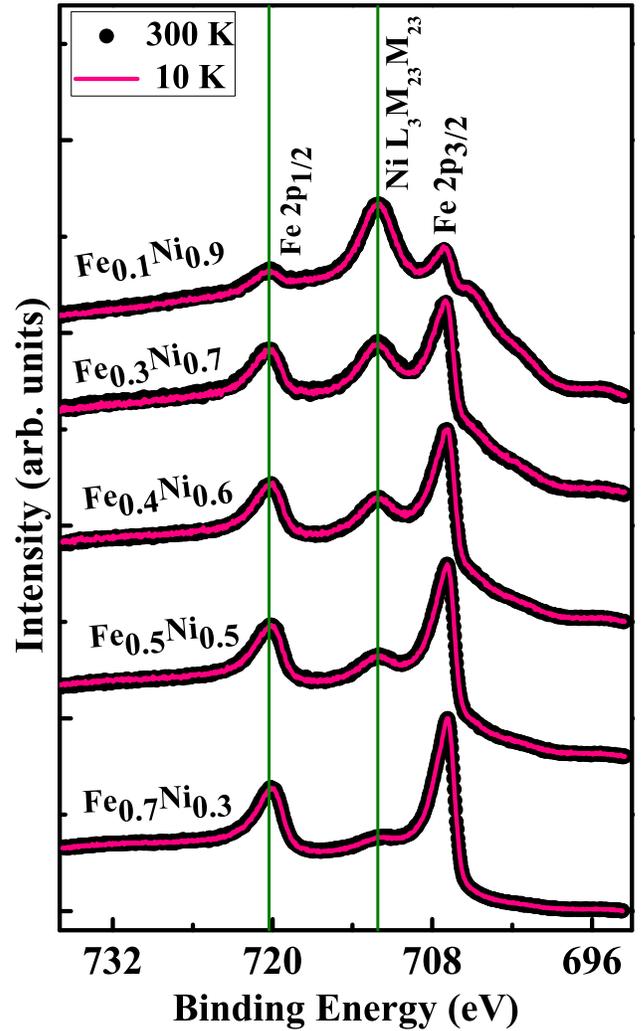}
 \vspace{-2ex}
\caption{Fe 2$p$ spectra of Fe$_{1-x}$Ni$_x$ for $x$ = 0.3, 0.5, 0.6, 0.7 and 0.9 recorded at 10 K (lines) and 300 K (symbols). The spectra at 10 K are overlayed on the spectra at 300 K. Sample composition is indicated above the spectra on the left side in the graph.}

\label{Fig7-Fe2pAll}
 \vspace{-2ex}
\end{figure}

The Fe 2$p$ spectra of Fe$_{1-x}$Ni$_x$ for $x$ = 0.3, 0.5, 0.6, 0.7 and 0.9 alloys are shown in Fig. \ref{Fig7-Fe2pAll}. As observed in Fig. \ref{Fig6-Fe2p}, the spectral functions recorded at 10 K and 300 K are almost identical to each other. In all these cases, the Fe 2$p$ signals are dominated by the well screened features as also observed in the case of pure Fe in Fig. \ref{Fig6-Fe2p}. The intensities at the higher binding energy side of 2$p_{3/2}$ signal can be accounted due to the presence of Ni $L_3M_{23}M_{23}$ Auger signal.

\begin{figure}
 \vspace{-2ex}
  \centering
\includegraphics [scale=0.4]{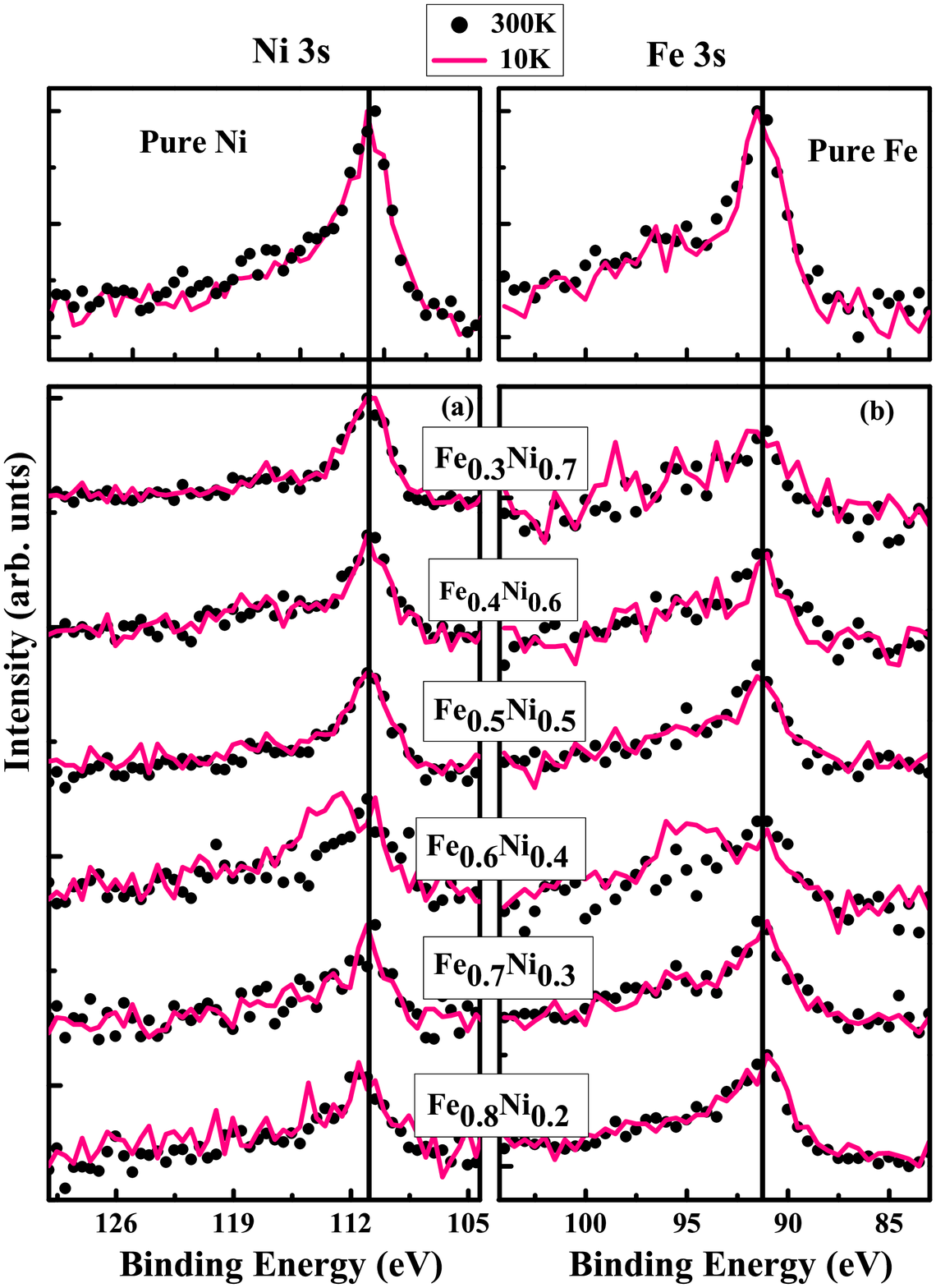}
 \vspace{-2ex}
\caption{(a) Ni 3$s$ and (b) Fe 3$s$ spectra of Fe$_{1-x}$Ni$_x$ recorded at 10 K (lines) and 300 K (symbols). The spectra at 10 K are superimposed on the spectra at 300 K. While all the spectra exhibit similar features at both the temperatures, the features in Fe$_{0.6}$Ni$_{0.4}$ are significantly different at the two temperatures. Top panels show Ni 3$s$ (left) and Fe 3$s$ (right) spectra taken on pure Ni and pure Fe metals respectively.}
\label{Fig8-Ni3sFe3s}
 \vspace{-2ex}
\end{figure}

We now turn to the shallow core level spectra, Ni 3$s$ and Fe 3$s$; the experimental results of Fe$_{1-x}$Ni$_x$ for various values of $x$ are shown in Fig. \ref{Fig8-Ni3sFe3s}(a) and Fig. \ref{Fig8-Ni3sFe3s}(b), respectively. The spectra recorded at 300~K and 10~K are superimposed for comparison. Both Ni 3$s$ and Fe 3$s$ spectra recorded at 300~K and 10~K temperatures match well for all the alloys except for the Fe$_{0.6}$Ni$_{0.4}$ Invar alloy. The spectra at 10~K exhibit an additional broad feature at high binding energy of about 2.6 eV and 3.8 eV away from the main peak in Ni 3$s$ and Fe 3$s$ spectra, respectively.

\begin{figure}
 \vspace{-2ex}
  \centering
\includegraphics [scale=0.4]{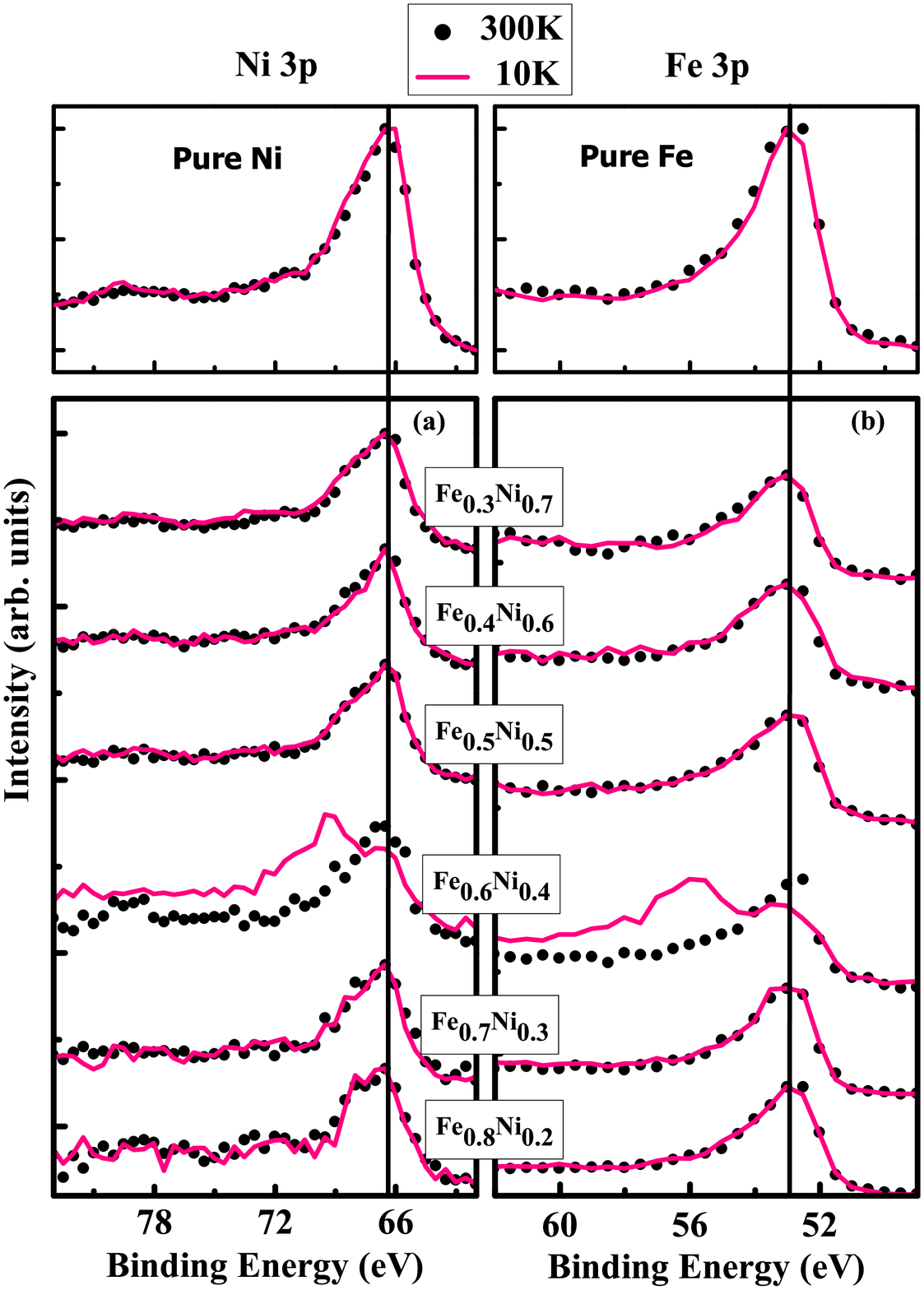}
 \vspace{-2ex}
\caption{(a) Ni 3$p$ and (b) Fe 3$p$ spectra of Fe$_{1-x}$Ni$_x$ recorded at 10 K (lines) and 300 K (symbols). The spectra at 10 K are superimposed on the spectra at 300 K. While all the spectra exhibit similar features at both the temperatures, the features in Fe$_{0.6}$Ni$_{0.4}$ are significantly different at the two temperatures. Top panels show Ni 3$p$ (left) and Fe 3$p$ (right) spectra taken on pure Ni and pure Fe metals respectively.}

\label{Fig9-Ni3pFe3p}
 \vspace{-2ex}
\end{figure}

In Fig. \ref{Fig9-Ni3pFe3p}, we show the Ni 3$p$ and Fe 3$p$ core level spectra for various compositions. The data recorded at 10~K and 300~K are superimposed for comparison. The intense broad signal consists of the spin-orbit split 3$p$ signal. Within the experimental detection level, we do not see signature of shift of the peak position as the changes are too small compared to the spectral width. Both Ni 3$p$ and Fe 3$p$ spectra recorded at 300~K and 10~K match well for all the alloys except for the Fe$_{0.6}$Ni$_{0.4}$ Invar alloy similar to the observations in the 3$s$ spectra. The 10 K spectra  of Fe$_{0.6}$Ni$_{0.4}$ exhibit an additional broad feature at higher binding energy of about 3.0 eV and 3.2 eV relative to the main peak for Ni 3$p$ and Fe 3$p$ spectra respectively; the energy separation seem to be larger than the separations derived in 3$s$ case.

\begin{figure}
 \vspace{-2ex}
   \centering
\includegraphics [scale=0.4]{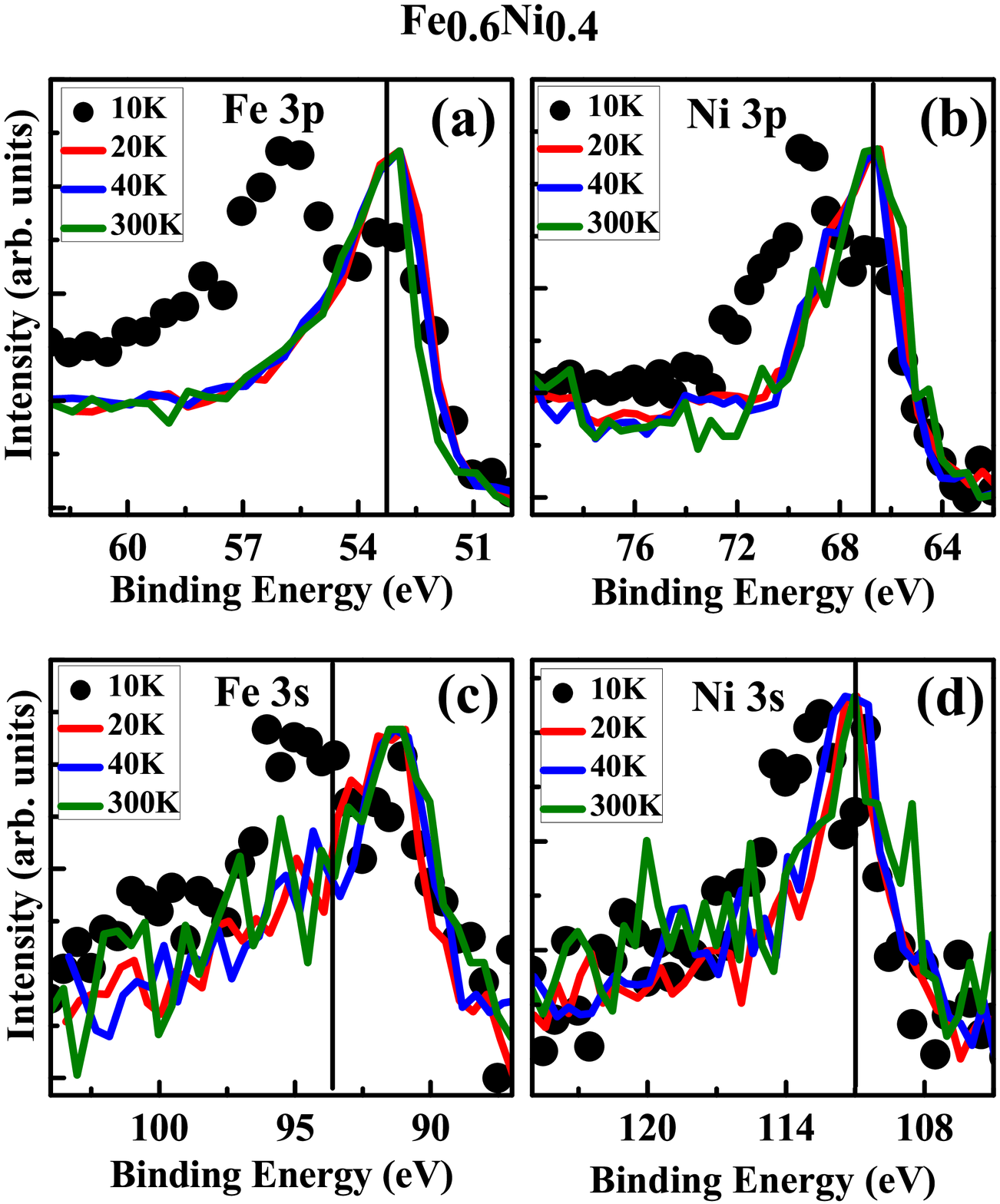}
 \vspace{-2ex}
\caption{Temperature evolution of (a) Fe 3$p$, (b) Ni 3$p$, (c) Fe 3$s$ and (d) Ni 3$s$ spectra of Fe$_{0.6}$Ni$_{0.4}$. All the spectra above 20 K exhibit similar linewidth. The data at 10 K exhibit an additional feature at higher binding energy side.}
\label{Fig10-3s3pTemp}
 \vspace{-2ex}
\end{figure}

In order to probe the temperature evolution of the spectral functions in the whole temperature range, the 3$p$ and 3$s$ spectra of Fe$_{0.6}$Ni$_{0.4}$ collected at different temperatures are investigated in Fig. \ref{Fig10-3s3pTemp}. The spectra recorded at 20~K, 40~K and 300~K match well with each other for both 3$p$ and 3$s$ spectra of Fe and Ni as also found in the deep core level spectra. Curiously, the spectra taken at 10~K exhibits a higher binding energy feature in every case shown here. In general, 3$p$ core level spectra of these 3$d$ transition metals are expected to be complex due to the presence of features arising from multiple effects such as electron correlation, small spin-orbit splitting and exchange splitting. Electron-electron Coulomb repulsion among the conduction electrons (electron correlation) often leads to multiple features in the photo-excitation spectra \cite{shakti,Swapnil}. There is a wide variation in the values reported for the exchange splitting of Fe 3$p$ core level; the reported exchange splitting varies from 0.26 eV to 1.1 eV \cite{campen,sinkovie,jungblut,sirotti,timura,rossi,roth}. The spin-orbit splitting was reported to be 0.7 eV \cite{timura} and 1.1 eV \cite{roth}. In that sense, 3$s$ spectra are relatively simpler due to the absence of spin-orbit effect. Still, we observe almost similar behavior in the 3$s$ spectra. Thus, the appearance of new feature in the 3$p$ and 3$s$ level at 10 K, signature of which has also been reflected in the deep core level spectra, indicate emergence of non-equivalent sites at low temperatures.

\begin{figure}
 \vspace{-2ex}
  \centering
\includegraphics [scale=0.4]{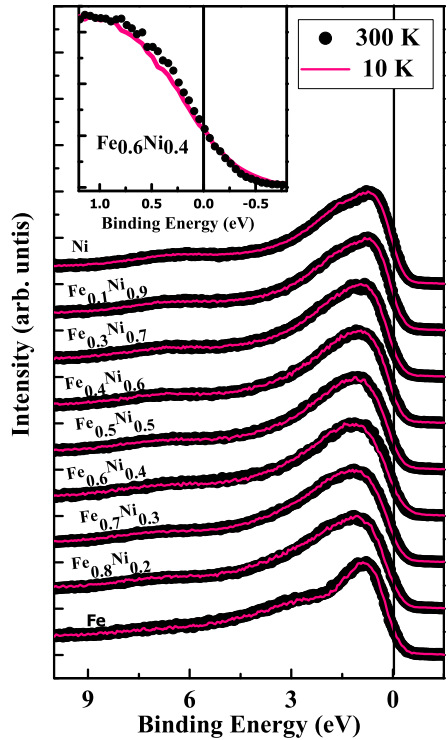}
 \vspace{-2ex}
\caption{Valence band spectra of Fe$_{1-x}$Ni$_x$ for various values of $x$ = 0.0, 0.2, 0.3, 0.4, 0.5, 0.6, 0.7, 0.9, 1.0. The spectra recorded at 10 K (lines) and 300 K (symbols) are superimposed for comparison. The inset shows the near Fermi energy region for Fe$_{0.6}$Ni$_{0.4}$ Invar alloy. The 10 K spectrum is appears to be broaden than 300 K spectrum!}
\label{Fig11-VB}
 \vspace{-2ex}
\end{figure}

In Fig. \ref{Fig11-VB}, we show the valence band spectra of Fe$_{1-x}$Ni$_x$ alloys collected at 300~K (symbols) and 10~K (lines). The valence band of Ni exhibits signature of three peaks. One intense peak near the Fermi level with maximum intensity at about 0.7 eV and a distinct shoulder at about 1.5 eV. The third peak appears at 6 eV binding energy. In Fe, the intense feature near Fermi level remains almost at the same energy but the shoulder feature appears at about 3 eV binding energy. The 6 eV binding energy feature becomes significantly weaker in intensity. The intermediate compositions exhibit a broad features in the vicinity of the Fermi level instead of two distinct features observed in the end members presumably due to broadening arising from the disorder induced effect.

No visible temperature dependence is observed in every case except the Invar alloy, Fe$_{0.6}$Ni$_{0.4}$. The relative intensities of the features appear to be somewhat different at the two temperatures. The 10 K spectrum of the Fe$_{0.6}$Ni$_{0.4}$ invar alloy becomes broader compared to that of 300 K spectrum suggesting, which is reflected by an anomalous spectral weight transfer to above Fermi level at lower temperature. This reveals signature of the formation of a mixed phase in the ground state presumably electronic in nature \cite{DDS-Phase}.

\section{Discussion}

The core level spectra of all the compositions of Fe-Ni alloys except the invar alloy exhibit no temperature dependence down to 10~K. In the invar alloy, Fe$_{0.6}$Ni$_{0.4}$, the deep core level (2$p$) spectra at 10 K becomes significantly more asymmetric towards the higher binding energy side than the spectra collected at 20 K and above. The outer core levels (3$s$, 3$p$) at 10~K exhibit signature of the emergence of an intense feature at higher binding energy side. The energy separation of the features developed at low temperature increases for outer core levels (3$s$, 3$p$) compared to the deep core levels (2$p$) presumably due to solid state effect of the shallow core levels \cite{Arindam-PRB}. The observed changes in the spectra take place below 20 K, where the Invar alloys are either in mixed magnetic state containing both ferromagnetic (FM) and antiferromagnetic (AFM) regions or in spin glass state \cite{rancourt1990}. The Fe-Ni invar alloys are inhomogeneous at microscopic level and possess three magnetically and crystallographically different phases. A reentrant temperature induced ferromagnetic spin glass transition was observed in Fe-Ni invar alloys \cite{miyazaki1985}. Previously, a three peak structure observed in Fe 3$p$ at 300~K for Fe$_{65}$Ni$_{35}$ alloy, which was attributed to the existence of two magnetic states of Fe at room temperature \cite{gawel2008,lomova2004,shabanova2001,shabanova1998,lomova2007}.

In Fe-Ni invar alloys, there are two high spin ferromagnetic phases with fcc ($\gamma$) and bcc ($\alpha$) structures having Fe magnetic moments of 2.8 $\mu_B$ and 2.2 $\mu_B$, respectively. A third fcc phase ($\gamma^\prime$) also appears with low spin configuration (magnetic moment of about 0.5 $\mu_B$), where Fe becomes antiferromagnetic below 20 K. The $\alpha$ and $\gamma^\prime$ phases get precipitated at low temperatures \cite{rancourtg,gawel2008}. In these alloys, a Fe site surrounded by small number of Fe atoms leads to low spin state of Fe and the Fe atom in the surrounding of ten to twelve Fe atoms takes high spin state. In the low spin state, Fe moment is compensated by surrounding Ni atoms. The core level spectra of Invar alloy in this study show no change in the temperature range 20 K - 300 K. Here, the Invar alloy is in fcc ($\gamma$) phase with Fe magnetic moment 2.8 $\mu_B$. But at 10 K, a drastic change occurs, which can be attributed to a large precipitation of antiferromagnetic, $\gamma^\prime$ phase, which is a low spin phase; in some regions, Fe gets surrounded by large number of Ni atoms leading to an additional feature at higher binding energy.

\section{Conclusion}

In summary, Fe$_{1-x}$Ni$_x$ ($x$ = 0.1, 0.2, 0.3, 0.4, 0.5, 0.6, 0.7 and 0.9) alloys were prepared by arc melting method and characterized by XRD, resistivity and magnetization measurements. While all the compositions exhibit soft ferromagnetic behavior, the invar alloy, Fe$_{0.6}$Ni$_{0.4}$ appears to be anomalous with a change in magnetization with temperature. Moreover, the invar composition exhibits highest negative magneto-resistance while both the end members possess positive magneto-resistance.

We studied the electronic structure of all the compositions employing $x$-ray photoelectron spectroscopic measurements. All the compositions except the invar alloy exhibit spectral features in line with those of the end members and did not exhibit temperature induced change in the electronic structure. The invar alloy, Fe$_{0.6}$Ni$_{0.4}$, however, exhibits anomalous temperature dependence in the core level spectra below 20~K. The marked temperature induced difference in the core levels of Invar alloy is attributed to a large precipitation of $\gamma^\prime$ phase in the ground state. Clearly, the ground state of invar alloy is quite complex and needs further investigation.

\section{Acknowledgements}

KM acknowledges financial support from the Department of Atomic Energy, Govt. of India under the project no. 12-R\&D-TFR-5.10-0100 and under the DAE-SRC-OI Award program, and Department of Science and Technology, Govt. of India under J. C. Bose Fellowship program.\\

VRRM acknowledges the extensive and useful discussion with Dr. W. Olovsson \& Prof. I. A. Abrikosov, IFM, Linkoping University on Fe-Ni Invars and Dr. Sunil Nair, IISER, Pune on magnetism and magneto-resistance of Fe-Ni alloys.\\

%

%
\end{document}